\documentclass[12pt]{aastex}

\newcommand{\Msun}{M$_{\odot}$\ }
\newcommand{\Lsun}{L$_{\odot}$\ }

\newcommand{\SI}{[SII]$\lambda$6731\ }

\newcommand{\Pa}{Pa$\beta$\ }
\newcommand{\Ha}{H$\alpha$\ }

\newcommand{\NII}{[NII]$\lambda$6583\ }

\newcommand{\km}{km s$^{-1}$\ }

\begin{document}


\title{The First X-shooter Observations of Jets from Young Stars$^{\star}$}


\author{F. Bacciotti\altaffilmark{1}}

\author{E.T. Whelan\altaffilmark{2}}

\author{J.M. Alcal\'a\altaffilmark{3}}

\author{B. Nisini\altaffilmark{4}}

\author{L. Podio\altaffilmark{5}}

\author{S. Randich\altaffilmark{1}}

\author{B. Stelzer\altaffilmark{6}}

\author{G. Cupani\altaffilmark{7}}

\altaffiltext{{}$^{\star}$}{Based on Observations collected with X-shooter at the 
VLT on Cerro Paranal (Chile), operated by the European Southern 
Observatory (ESO). Program ID: 085.C-0238(A)}
\altaffiltext{1}{INAF - Osservatorio Astrofisico di Arcetri, Largo E. Fermi
5, 50125 Firenze, Italy}
\altaffiltext{2}{Laboratoire d'Astrophysique de
l'Observatoire de Grenoble, UMR 5521 du CNRS, 38041 Grenoble Cedex,
France}
\altaffiltext{3}{INAF - Osservatorio di Capodimonte, Via Moiariello 16, 80131
Napoli, Italy}
\altaffiltext{4}{INAF - Osservatorio di Roma, Via di
Frascati 33, 00040, Monteporzio Catone, Italy}
\altaffiltext{5}{Kapteyn Astronomical Institute, Landleven 12, 9747 
AD Groningen, The Netherlands}
\altaffiltext{6}{INAF-Osservatorio di Palermo, Piazza del Parlamento 1, 90134 Palermo, Italy}
\altaffiltext{7}{INAF-Osservatorio di Trieste, Via Tiepolo 11, 34143 Trieste, Italy}



\begin{abstract}

We present the first pilot study of  
jets from young stars conducted with X-shooter, on ESO/VLT. 
As it offers simultaneous, high quality spectra in 
the range 300-2500 nm X-shooter 
is uniquely important for spectral diagnostics in jet studies. 
We chose to probe 
the  accretion/ejection mechanisms at low stellar masses examining 
two targets with well resolved continuous jets lying on the plane of the sky,
ESO-HA 574 in Chamaleon I, and Par-Lup3-4 in Lupus III. The mass of the latter is close to 
the sub-stellar boundary (M$_{\star}$=0.13 M$_{\odot}$).

A large number of emission lines probing  
regions of different excitation are identified, position-velocity diagrams are presented and 
mass outflow/accretion rates  are estimated. Comparison between the 
two objects is striking.
ESO-HA 574 is a weakly accreting star 
for which we
estimate a mass accretion rate of $log(\dot{M}_{acc}) = -10.8 \pm 0.5$ 
(in \Msun yr$^{-1}$), 
yet it drives a powerful jet with 
$\dot{M}_{out}$\,$\sim 1.5-2.7 \times 10^{-9}$ \Msun yr$^{-1}$.
These values can be reconciled with a magneto-centrifugal 
jet acceleration mechanism only assuming that the presence 
of the 
edge-on disk severely depresses the luminosity of the  accretion tracers. 
In comparison Par-Lup3-4 with 
stronger mass accretion 
($log(\dot{M}_{acc}) = -9.1 \pm 0.4$ \Msun yr$^{-1}$), 
drives a low excitation jet with about
$\dot{M}_{out}$\,$\sim 3.2 \times 10^{-10}$ \Msun yr$^{-1}$ in both lobes.
Despite the low stellar mass, $\dot{M}_{out}$/$\dot{M}_{acc}$ for Par-Lup3-4 is  
at the upper limit of the range usually measured for young objects,  but   
still compatible with a steady magneto-centrifugal wind scenario if 
all uncertainties are considered.
\end{abstract}





\section{Introduction}

Collimated outflows are common to accreting objects with 
a wide variety of spatial scales and
mass accretion rates, but the phenomenon is best 
studied in the case of jets from the nearby     
low-mass stars in their pre-main sequence phase \citep{Ray07, Whelan09}. 
These flows present  characteristic 
spectra rich in shock-excited emission lines
\citep{Ray07, HMR94},
that allow a detailed diagnostic of the kinematic
and of the physical properties of the gas. 
To  better understand the 
enigmatic process of jet generation,  
it is imperative to determine the exact relationship between
outflows and accretion. Their interplay is at the
heart of the widely accepted theory of magneto-centrifugal launching,
that may be common to all jet generating objects 
\citep{Ferreira06, Pudritz07,Shang07}. The 
pertinent models prescribe stringent constraints for the ratio
of accreted to ejected mass outflow rates, 
and it is important to understand how this relationship 
changes with the mass of the driving sources, and in particular
for low stellar masses
\citep{Whelan10}.

The spectra of stellar jets 
can today be efficiently analysed with dedicated diagnostics of key lines, and 
estimates of mass outflow rates ($\dot{M}_{out}$) 
can be compared with corresponding mass accretion rates 
($\dot{M}_{acc}$) \citep{BE99, masc04, Podio06, Hartigan07}. Reliable estimates of 
$\dot{M}_{out}$/$\dot{M}_{acc}$, however, can only be given if 
one has a detailed knowledge of the excitation properties 
of the gas, and in each location. It is thus critical
to have a complete spectral mapping along the jet extension,  
in the largest possible wavelength range.

With these ideas in mind we chose to characterise the jets of ESO-HA 574 and Par-Lup3-4,
two intriguing objects 
included in a  pilot study of young stars of low and sub-stellar mass  with X-shooter.
Mounted on the Unit
Telescope 2 of the Very Large Telescope (VLT), X-shooter 
is an instrument 
providing high quality, medium resolution, UBV (300-559.5 nm), VIS
(559.5-1024 nm) and NIR (1024-2480 nm) spectra.  This wavelength range
includes a large number of emission lines typical of both outflow and
accretion phenomena, making X-shooter currently unequaled in
spectral diagnostic studies.  

The outflows from the selected sources have a continuous jet-like 
morphology and lie close to the plane of the sky, which makes them   
well resolved spatially with no superposition of the blue 
and red lobes, and with minimum obscuration by the disk. 

Our first target is ESO-HA 574 
(11$^{h}$16$^{m}$03$^{s}$7, -76$^{\circ}$24$^{\arcmin}$53$^{\arcsec}$),
a low-luminosity, low-mass  source discovered by 
\cite{Comeron04} in a study of the Chamaleon I star forming 
region (d=160 $\pm$ 17 pc, \citealt{whi98}). Its bolometric 
luminosity is only 0.0034 \Lsun \citep{Luh07}, that, compared  
with other typical T Tauri stars of the same spectral type (K8) 
in Chamaleon I, makes this star underluminous by a factor  
of about 150. This  
is probably caused by the edge-on viewing geometry of its disk.
The lines normally used for accretion
diagnostic are also very weak \citep{Comeron04}, 
which may suggest low levels of accretion, 
yet the star powers a well developed bipolar jet (HH 872, of total length 
of 3150 AU) at  position angle (PA) $\sim$ 45$^{\circ}$, revealed in
follow-up FORS1 [SII] imaging by \cite{Comeron06}. The jet is made
up of 5 distinct knots, with knots A, B, C, D forming the
blue-shifted jet, and knot E in the red-shifted lobe. 

The second target, Par-Lup3-4 (16$^{h}$08$^{m}$51$^{s}$44,
-39$^{\circ}$05$^{\arcmin}$30.5$^{\arcsec}$), was first reported by
\cite{Nakajima00} as part of a near-infrared survey of the Lupus III
dark cloud (d=200 $\pm$ 40 pc, \citealt{Comeron03}). 
Its mass is only $\sim 0.13$  M$_{\odot}$, but 
the accretion tracers in the spectrum are quite strong, and indicate a 
mass accretion rate of  $\dot{M}_{acc} = $ 1.4~10$^{-9}$  M$_{\odot}$ yr$^{-1}$ 
\citep{Comeron03}. 
The forbidden lines are associated with a small collimated jet 
(HH 600) at PA $\sim$
130$^{\circ}$ \citep{Comeron05, Comeron11}. 
The source is under-luminous by a factor of about 25 with respect to other 
M5 young stars in Lupus III, 
its bolometric luminosity being 0.003\,L$_\odot$ \citep{mer08}. 
An edge-on disk oriented at about 81$^{\circ}$  is probably responsible for 
the subluminosity \citep{Hue10}. 

We extend these studies, that were limited to the visual band,
by taking X-shooter spectra of the targets 
with the slit parallel to the jet axis. 
We thus obtained  
in a single observation, and for the first time in the field
of stellar jet studies, a simultaneous, complete spectral 
mapping from 300 nm to 2.2 $\mu$m,
that provided 
immediate and far-reaching information about the
physical properties of the outflows and of their  
sources. 
In this letter we present the spectra and discuss the accretion/outflow 
properties of the two targets.  A complete mapping of the jet physical 
conditions derived from  the same datasets  is the subject of a second paper in preparation.

\section{Observations and Data Reduction}

The observations  
were conducted on April 7th 2010 (Program ID: 085.C-0238(A)), 
as part of the INAF/GTO program 
on star forming regions \citep{alcala11}. 
The seeing varied between 0\farcs9 and 1\farcs2.
In both cases the 11$''$-long slit 
was aligned with the  jet axis, along position angle (P.A.)
44.8 and 130.0 degrees for ESO-HA 574 and Par-Lup3-4, respectively (see above). 
The total exposure time per star was 3600 s (4$\times$900) in all arms  
and nodding observing mode ``ABBA" was used (shifting along the jet PA) 
in order to minimize the contribution of the background in the NIR.
Slit-widths of 1\farcs0, 0\farcs9 and 0\farcs9 were used, 
yielding resolving powers of 5100, 8800 and 5600 (corresponding to about 58, 34
and 53 \km velocity resolution) for the UBV, VIS and NIR arms, respectively. 
The data reduction was performed independently for each arm using 
the X-shooter pipeline version 1.1.0 \citep{mod10}. 
One-dimensional spectra were then extracted over the star continuum and over the 
extension of the knots.
Note that the latters appear  
only in either A or B nodding position, so the 
spectra over each knot correspond to an integration time of 1800 s.  
The reported radial velocities are systemic, with the LSR velocity of the sources 
determined using the photospheric absorption line of KI\,$\lambda$7698 (VIS arm).
For ESO-HA 574 we  measured $v_{LSR}=$ +3  km s$^{-1}$, while for Par-Lup 3-4 
$v_{LSR}=$ 5.5 km s$^{-1}$, 
in agreement with the LSR velocities 
of the corresponding clouds, measured from CO observations to be 
4  and 7  $\pm 2$ km s$^{-1}$, respectively \citep{dame87}.
On-source spectra were extinction corrected 
assuming for ESO-HA 574 A$_{v}$=1.7 mag (derived from A$_J$=0.45 
\citep{Luh07}, using the extinction curve at R$_v$=3.1 from \cite{sav79}), 
and A$_{v}$=3.5 mag for Par-Lup3-4 \citep{Hue10, Comeron04}. 
Nevertheless, for both targets A$_v$ values around zero are derived off-source 
using the [FeII] lines at 1.27, 1.32, 1.64 $\mu$m, 
that trace the jet emission  \citep{Nisini05}. This suggests that 
the jet propagates out of the envelope obscuring the star, and 
therefore the spectra extracted for the jet knots were not dereddened.

\section{Results}

\begin{figure}
 \includegraphics[width=22cm, angle=-90]{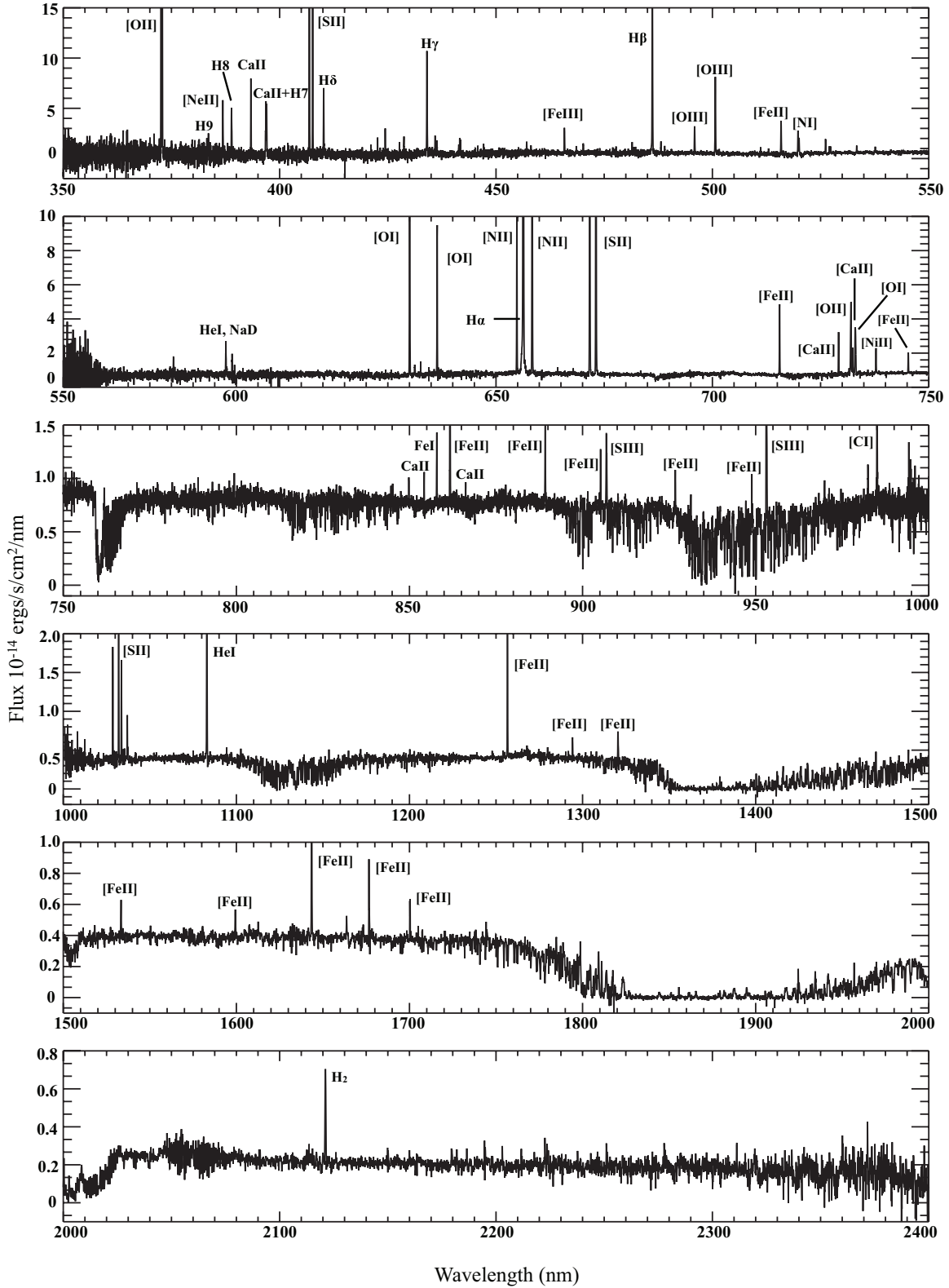}
  \caption{Full extinction corrected (A$_{v}$ = 1.7) 
X-shooter spectrum of ESO-HA 574, integrated over $\sim$ 1$''$ around 
the source.}  
  \label{esospec}
\end{figure}

Figure \ref{esospec} shows the X-shooter spectrum of ESO-HA 574 extracted at 
the source position, obtained by summing over 5 to 7 pixel rows 
(0\farcs9 to 1\farcs1, depending on the PSF of the considered arm). 
The system is observed in numerous lines, including both accretion and ejection tracers.  
Balmer lines from H9 to H$\alpha$ are identified, as well as the calcium triplet 
and He\,I at 1.083 $\mu$m and Pa$\beta$, while Br$\gamma$ is not detected. 
Among the many forbidden lines, we find strong lines of sulfur, 
oxygen, nitrogen and iron in multiple ionisation states.
As noted by \cite{Comeron04}, the forbidden [CaII]$\lambda$7321,7324 lines 
are much more prominent than the calcium triplet, supporting the idea that 
both the stellar continuum and the accretion tracers  are occulted by the edge-on disk,
while the jet tracers are unaffected (as it occurs in the edge-on system T Cha studied by \cite{schi09}).

In Figure \ref{pvs} we present Position-Velocity (PV) diagrams of a selection of key lines. 
Knots A, B (jet) and  knot E (counter-jet)
are indicated in the [SII]$\lambda$6731 panel. 
Knot A1, identified in \cite{Comeron06} as the source, has actually moved 
to about 0\farcs2 with respect to the star continuum. 
The emission in H$\alpha$ and H$\beta$, as well as in 
[OII]$\lambda\lambda$3726,3729 and \SI\, lines 
is  extended, with higher dispersion close to the star 
and in knot E, that has a bow-shock shape. 
On the contrary,
the emission in the H$_2$ lines is 
concentrated around the star, as often in Classical T Tauri stars (CTTSs) \citep{Beck08},
being likely originated in shocked walls of a cavity opened by the flow.  
In general the lines present low systemic radial velocities, with knots A and B  
blueshifted at  $\sim$ -3 and $\sim$ -6
\km in \SI\,, while knot E is redshifted at +25 \km.
Tangential velocities were determined by comparison with the [SII] images of
\cite{Comeron06}, taken on 21 July 2005. We find
V$_{tan} =$ 130, 220 and 325 km s$^{-1}$ for knots A, B and E, respectively. 
Large knot velocity differences and velocity asymmetries in 
opposed jet lobes are commonly observed in CTTS jets 
\citep{Hirth94, Podio11}.
The determined velocities suggest very low jet 
inclination angles of only 1.3$^{\circ}$, 1.6$^{\circ}$ and 4$^{\circ}$ 
at knots A, B and E, respectively, 
confirming that the disk is nearly edge-on, and it is causing the
observed under-luminosity \citep{Comeron04}.

Turning to Par-Lup3-4, the full spectrum on-source 
(integrated over the 1$''$ of the continuum width) is shown in Figure \ref{parlupspec}, 
and three selected PV diagrams in Fig. \ref{pvs} (bottom). 
Far fewer jet tracers are detected than in ESO-HA 574, 
while accretion signatures like the calcium triplet at 8500 \AA, are stronger. 
\Pa and Br$\gamma$ are both present here.
Only a few lines are extended, notably the [SII] lines, that show   
a well-developed bipolar structure, extending to $\sim$2\farcs5 (500 AU) on both sides. 
Radial velocities up to -20 and +20 kms$^{-1}$ are measured, 
consistently with the LSR velocities of  -18.3 and +23.6 km s$^{-1}$ determined by 
\cite{Comeron05} in the same line. 
The strong [OI]$\lambda$6300 line is also extended, although the red-shifted
component is affected by subtraction of the [OI] atmospheric emission 
(see inset in Figure \ref{pvs}).  The weak \NII shows also traces of bipolarity, 
over similar radial velocities.
Tangential velocities have been reported recently by \cite{Comeron11} to 
be 170 $\pm$ 30 \km in both jet lobes, that combined with our radial velocities give a 
jet inclination angle of about 7$^{\circ}$.

\section{Discussion} 

A large number of accretion tracers are 
identified in the spectra at the star position. 
We estimated mass accretion rates as in \cite{alcala11}, 
using the relationship 
$\dot{M}_{acc} \approx 1.25 \cdot L_{acc} R_{\star}/G M_{\star}$ 
where $R_{\star}$ and $M_{\star}$ are the stellar 
radius and mass, respectively, and  $L_{acc}$  
the accretion luminosity \citep{hartmann98}. 
The latter is obtained from the luminosities of the different lines, 
via the relationships in 
\cite{Herczeg08}, \cite{mohanty05}, \cite{muzerolle98}.  
The line luminosities are derived from the  extinction-corrected absolute fluxes
obtained  by simple integration in the calibrated spectra.
Adopted stellar masses and radii, estimated from stellar luminosity and 
spectral type (see Sect.\ 1) are 0.5\,M$_\odot$ and 0.2\,R$_\odot$ 
for ESO-HA~574 and 0.13\,M$_\odot$, and 0.45\,R$_\odot$ for Par-Lup~3-4. 
The results are reported graphically in the top panels of 
Fig. \ref{accretion}, where 
the error on each point derives from the uncertainty in the  
calibration relationships 
and from the noise level of the continuum. 
We considered accretion tracers in all the available wavelength domain, including 
Pa${\beta}$ and Br${\gamma}$ (illustrated in middle panels of Fig. \ref{accretion}), while
the Balmer jump could not be investigated because of the low 
signal-to-noise in the bluest part of the spectra. 
Excluding [OI], which is dominated by the wind, 
the determined  values for  $\log{\dot{M}_{acc}}$ are fairly consistent within 
the errors.  The absence of a wavelength dependence is in line with the attenuation caused by
an edge-on disk, as the dust extinction from large grains in the disk is not expected to vary in wavelength 
\citep{schi09}. 
Excluding [OI], the values give an  average of $-$10.8$\pm$0.5 
and $-$9.1$\pm$0.4 (in M$_\odot$ yr$^{-1}$), for ESO-HA~574 and Par-Lup~3-4, 
respectively, where the indetermination is the standard deviation.  
The $\dot{M}_{acc}$ value for Par-Lup~3-4 is in good agreement with the 
one derived by \citet{Comeron03}. The extremely low value for 
ESO-HA~574, however, can hardly be reconciled with the presence of a 
powerful jet, and the estimate is probably severely affected by the 
obscuration caused by the edge-on disk. Indeed, assuming that both 
the continuum and the accretion tracers are depressed by the same amount, 
the correction for the disk occultation on $L_{\star}$, and, consequently, on 
$R_{\star}$ and $L_{acc}$, would bring to a mass accretion rate of 
$\sim$1.7~10$^{-8}$\,M$_\odot$ yr$^{-1}$, in line with other stars of 
similar mass \citep{natta04}. 
On the contrary, for Par-Lup3-4 $\dot{M}_{acc}$
is just above the typical value for similar stars, 
suggesting that the 81$^{\circ}$ inclination of the disk plane \citep{Hue10} 
leaves the accretion funnels visible at least in part.

A comparison of the H$\alpha$ profile in the two targets 
is given in Fig. \ref{accretion},  bottom.
For ESO-HA 574 the majority of the H$\alpha$ emission is extended and relatively 
narrow (-100, +100 \km), superimposed to a much weaker broad component (-300, +300 \km), 
localized at the star continuum. 
On the contrary in Par-Lup3-4 the H$\alpha$ line is concentrated around the star, 
it is broad (10$\%$ width $\sim$ 350 \km), and its shape is normally not associated to a jet 
component (type II B profile in the classification of \citealt{reip96}). 
In order to disentangle the emission pertinent to accretion or to 
ejection, the \Ha line profiles were analysed with 
spectro-astrometry a technique that  
allows one to find positional displacements of 
the line peak with respect to the star continuum  
over angular distances of a few tens of milliarcseconds 
(for details see \citealt{whegar08}). 
The absence of any significant offset for Par-Lup 3-4 
is consistent with the disappearance of the jet in \Ha in the 2010 images of 
\cite{Comeron11}, and suggests an accretion origin for this line.
The same applies for the broad \Ha component in ESO-HA 574, while 
the jet portion between $\pm$ 100 \km has obviously a large offset, 
preventing the use of the accretion diagnostic based on the width
at 10\% intensity \citep{natta04}.

\begin{figure}
 \includegraphics[width=20cm, angle=-90]{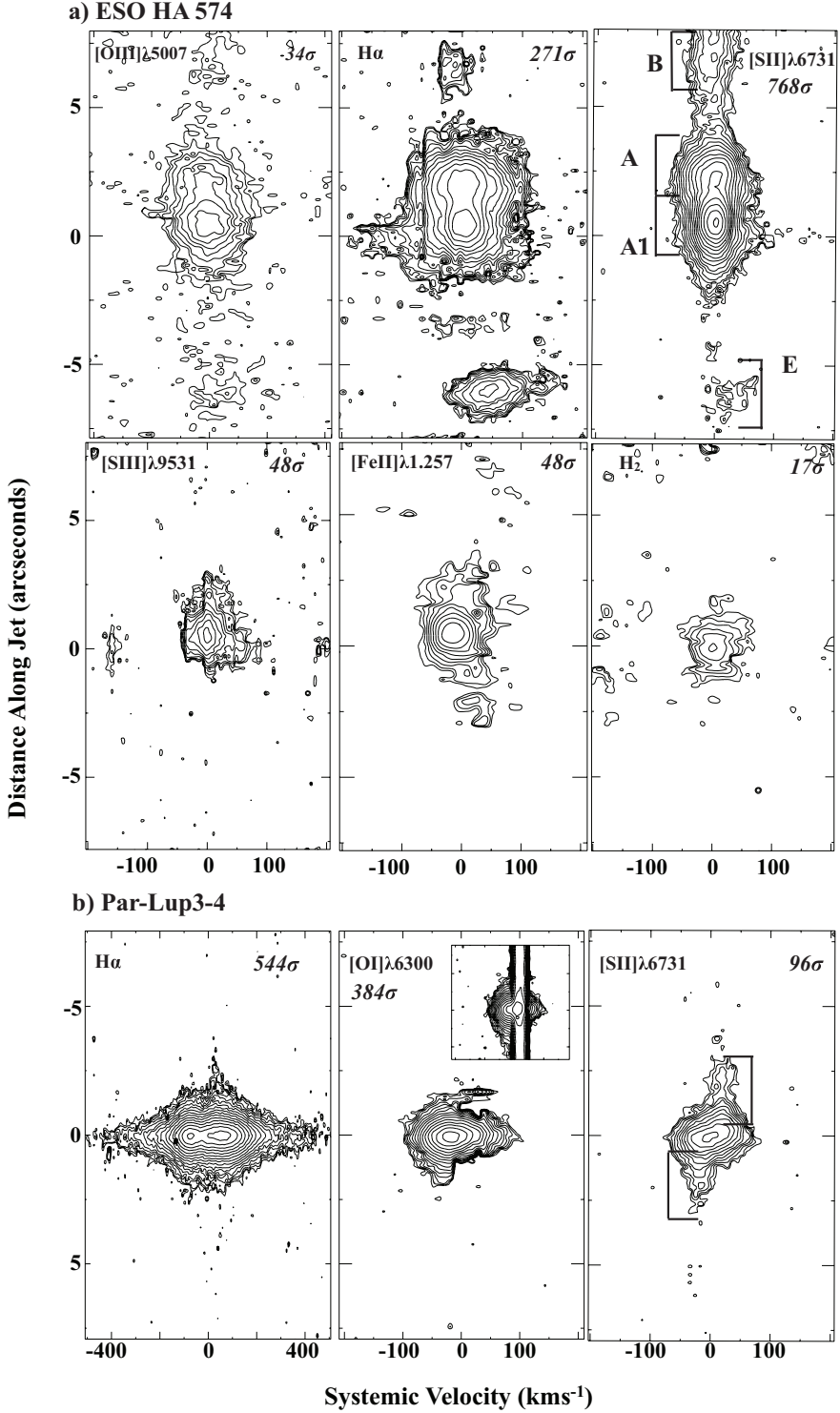}
    \caption{Position-Velocity (PV) diagrams 
of selected lines around ESO-HA 574 and Par-Lup 3-4. 
Contours begin at 3 times the r.m.s. noise and increase by a factor 
of $\sqrt{2}$. The peak value is shown in the upper right corners. 
1D spectra of the knots in the ESO-HA 574 jet and the 
two lobes of the Par-Lup3-4 jets  were extracted over the regions 
marked on the [SII]$\lambda$6731 panels. 
Distances along the blue-shifted lobe have positive $y$ values.
}
\label{pvs}
\end{figure}

\begin{figure*}
 \includegraphics[width=22cm, angle=-90]{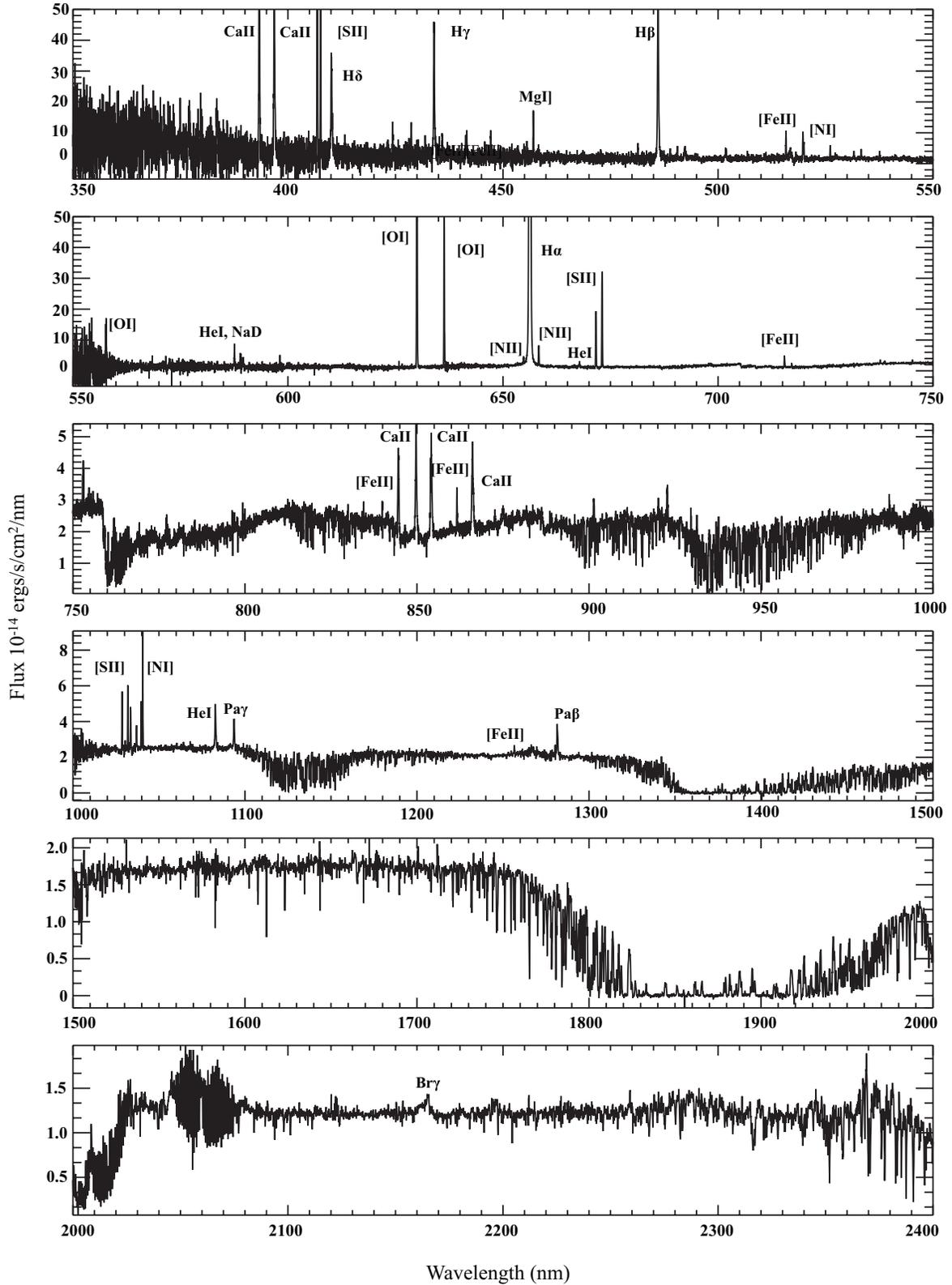}
  \caption{Same as Fig 1, for Par-Lup3-4. The spectrum is  extinction-corrected for  
A$_{v}$=3.5.}
\label{parlupspec}
\end{figure*}

\begin{figure}
 \includegraphics[width=20cm, angle=-90]{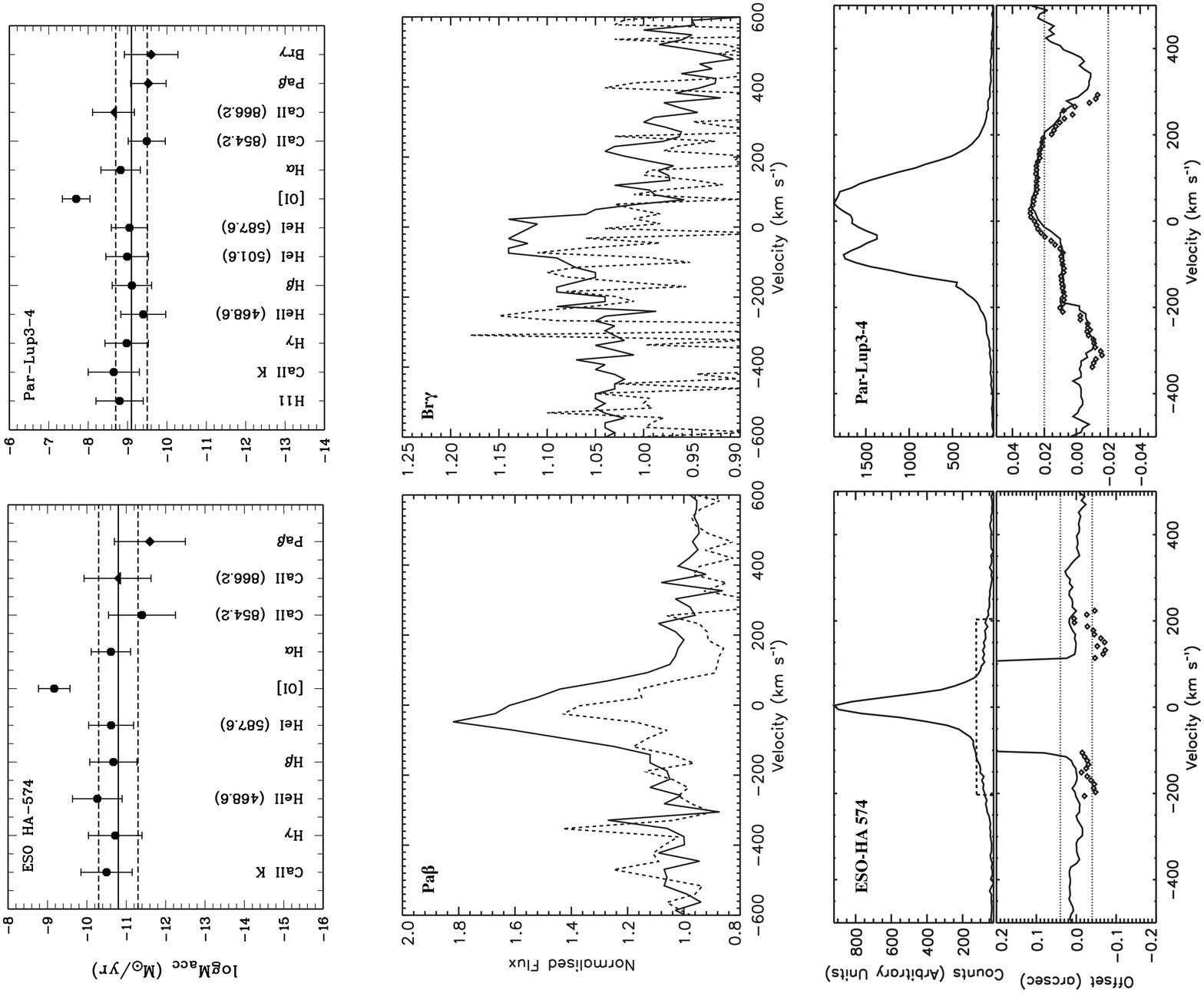}
    \caption{
{\em Top}: 
Mass accretion rates from different accretion tracers. 
Solid line: average (excluding [OI]); dashed line: $\pm 1\sigma$ standard deviation.
{\em Middle:} Pa$\beta$ and Br$\gamma$ in  
ESO-HA 574 (dashed line) and Par-Lup3-4 (solid line), normalised to their continua. 
{\em Bottom}: spectro-astrometric signal in \Ha, before (solid curve) and after 
(diamonds) continuum subtraction.
Positive offsets are along the blue-shifted jet
(dashed lines: spectro-astrometric accuracy). 
}
\label{accretion}
\end{figure}

Regarding the jets, X-shooter fully reveals  its power showing at once  
the actual excitation state of the gas along the flow. 
Fig. \ref{pvs} shows that in ESO-HA 574 strong [OIII]$\lambda$5007 
and [SIII]$\lambda$9531 lines are observed. The 
knots closer to the source (A1, A) present the highest excitation and density 
as also probed by [SII]$\lambda$4069,
\NII, and by the [SII] and [NI] lines at 1.03/1.04 $\mu$m.
The emission in the helium line at 1.083 $\mu$m  is also very strong, 
and spatially extended, as it has been found in other CTTSs  \citep{Takami02, Podio08}. 
This line is a probe of inner wind regions \citep{kwa07}, and  indicates
high electron temperatures ($T_e \sim$ 10$^4$ K) and densities 
($n_e > 10^6$ cm$^{-3}$). In the same way in Par-Lup3-4 
the absence of lines from high ionisation states, and the 
strength of  [OI] and [NI] lines with respect to [NII] lines
reveal that the gas in this jet has a low excitation state all along the flow.

Importantly, this qualitative inspection indicates that the usual diagnostic procedures 
to estimate  mass outflow rates \citep{BE99, Podio06} 
cannot be applied for ESO-HA 574, as the abundances 
of the species considered (OI, FeII, NII) should be calculated  
considering high ionisation levels. 
On the other hand in the Par-Lup 3-4 jet, instrinsically of low excitation, 
[NII] lines are extremely faint and [OI] lines 
are contaminated by the sky lines.  
Therefore here we give only a preliminar estimate of
the mass outflow rates ($\dot{M}_{out}$) from  
the [SII]$\lambda$6731 line luminosity, in the knots where [SIII] lines are not detected. 
Following  the method in \cite{Hartigan95}, in each knot $\dot{M}_{out}$ is calculated as: 
\begin{equation}
\dot{M}_{out}~~({\rm M}_{\odot} ~{\rm yr}^{-1}) = 4.51 \times 10^{9}~ 
(1 + \frac{n_{c}}{n_{e}})~  
{\rm L}_{{\rm [SII]}\lambda6731} ({\rm L}_{\odot})~ 
{ V_{tan} ({\rm km ~s}^{-1}) \over l_{tan}({\rm cm})}
\label{eq1}
\end{equation}
where n$_{c}$ is the critical density for the [SII] ratio
(2~10$^4$ cm$^{-3}$),  while V$_{tan}$ and l$_{tan}$ are 
the tangential velocity and the knot length, respectively.
The line ratios and luminosities 
were estimated from 1D spectra extracted  
over the spatial regions marked in Figure 2 ([SII]$\lambda$6731 panel), with no
correction for extinction (see Sect. 2). 
The errors are propagated from the noise level around
the [SII] lines, and from the distance to the star.
  
The results are collected in Table \ref{table1}. 
For ESO-HA 574 $n_e$ turns out to be 1750, 280 and 110 cm$^{-3}$ for knots 
A, B and E respectively, well below the critical density.
In knot A1 $n_e = 3000 \pm$ 130  cm$^{-3}$, but  $\dot{M}_{out}$
could not be estimated because [SIII] lines are detected here, while the method
assumes sulfur to be all singly ionised. In the other knots  
$\dot{M}_{out}$ turns out to be  1.5 - 3 10$^{-9}$ M$_{\odot}$ yr$^{-1}$,  
within the usual range found for  CTTS jets
\citep{Hartigan95, meln09}. Notably, 
values are similar along the flow, as expected if no big sideway losses are present 
(due, e.g. to large bow-shocks), 
and, as in other asymmetric jets,  they are about the same in the two lobes, 
despite the gas  
has different physical conditions (see \citealt{Podio11} for a discussion). 
The obtained $\dot{M}_{out}$ however, appears two orders of magnitude higher than the
derived  $\dot{M}_{acc}$, if the latter is taken at face value. As discussed above, 
however, the occultation of the edge-on disk should be taken into account. 
A correction of $\dot{M}_{acc}$ for the observed under-luminosity would bring 
the mass ejection/accretion ratio back 
to $\approx$ 0.3 (over the two lobes), in the expected 
range for  magneto-centrifugal jet launch
(0.01 $< \dot{M}_{out} / \dot{M}_{acc} <$ 0.5, \citealt{Cabrit09}).  

For Par-Lup 3-4, $\dot{M}_{out}$ was estimated 
separately for each lobe, 
coadding the spectra over the two regions shown in Figure \ref{pvs}, bottom panels.
The [SII] line ratio provides here  $n_e =$ 1150 
and 3200 cm$^{-3}$, respectively. The [SII]$\lambda$6731 luminosity was  
evaluated over the same regions, and we take $v_{tan} = 170 \pm 30$ 
kms$^{-1}$ in both lobes \citep{Comeron11}. Using Eq.\,\ref{eq1} we 
obtain $\dot{M}_{out} =$ 3.6 10$^{-10}$ M$_{\odot}$ 
yr$^{-1}$ for the blue jet and 3.2 10$^{-10}$ M$_{\odot}$ yr$^{-1}$ for the red lobe,
remarkably similar on the two sides of the system.
Therefore, over the two lobes  $\dot{M}_{out}/\dot{M}_{acc} \sim 0.85$,
which is slightly above the maximum indicated for
magneto-centrifugal acceleration. The estimated uncertainties on 
$\dot{M}_{acc}$ and  $\dot{M}_{out}$, however, 
combined with the partial obscuration of the accretion 
funnels by the close-to-edge-on disk
can lower this value to 0.3 - 0.5, compatible with that scenario.


Summarising, 
X-shooter spectra prove to be fundamental in jet studies 
as they provide an immediate overview of the outflow kinematics and excitation,
as well as a rich accretion diagnostics.  
The first two low-mass (0.5 and 0.13 M$_{\odot}$) targets examined proved to 
have accretion/ejection properties compatible with a 
magneto-centrifugal scenario. 
An extension of the current diagnostic techniques 
to fully exploit the potential of X-shooter spectra 
is underway and will be presented in a forthcoming paper.

\acknowledgements{We dedicate this letter to the memory of 
Roberto Pallavicini. We thank V. D'Odorico, P. Goldoni, A. Modigliani 
and the ESO staff for support with the observations and the X-shooter pipeline,
and the referee for very useful comments.   
E.T. Whelan is supported by an IRCSET-Marie Curie Fellowship 
in Science within the European Community FP7.}

\begin{table}
\begin{center}
\begin{tabular}{cccccc}
\hline
\hline   
Object &n$_{e}$   &L$_{{\rm [SII]}\lambda6731}$   &V$_{tan}$  &l$_{tan}$  &$\dot{M}_{out}$  \\
 &(10$^3$ cm$^{-3}$)   &(10$^{-7}$~\Lsun)  &(km s$^{-1}$) &(10$^{15}$ cm) &(\Msun yr$^{-1}$) \\
\hline
ESO HA 574  & & & & &
\\
Knot A            &1.7 $\pm$ 0.1   &7.7 $\pm$ 0.5  
    &130 $\pm$ 30 &3.8 $\pm$ 0.4 & 1.5 $\pm$ 0.4  
$\times$ 10$^{-9}$
\\
Knot B &0.28  $\pm$ 0.03       &2.4 $\pm$ 0.4        
&220 $\pm$ 50   &6.5 $\pm$ 0.7 &2.7 $\pm$ 1  
$\times$ 10$^{-9}$
\\
Knot E &0.11   $\pm$ 0.1      &0.42 $\pm$ 0.07      &325  
$\pm$ 50 &5.3 $\pm$ 0.6  &2.5 $\pm$ 2 $\times$  
10$^{-9}$
  \\
  Par-Lup3-4 & & & & &
   \\
  Blue lobe &1150 $\pm$ 20 &1.85 $\pm$ 0.3  &170 $\pm$ 30  
&7.2 $\pm$ 1.5 &3.6 $\pm$ 0.6 $\times$ 10$^{-10}$
  \\
Red lobe &3200 $\pm$ 50 &5.2 $\pm$ 0.5  &170 $\pm$ 30 &9.0  
$\pm$ 1.8 &3.2 $\pm$ 0.3 $\times$ 10$^{-10}$
  \\
\hline
\end{tabular}
\caption{Mass outflow rates in the jets from 
ESO-HA 574 and Par-Lup3-4, estimated from the luminosity of  
the [SII]$\lambda$6731 line (see text). 
}
\label{table1}
\end{center}
\end{table}
{}
\end{document}